# STRUCTURAL FEATURES OF QUASICRYSTALLINE NANOPARTICLES WITH SEVENTH-ORDER ROTATIONAL SYMMETRY

Alexander S. Prokhoda, PhD

**Ukraine**

a-prokhoda@ukr.net

In this study, we present atomic models of nanoparticles exhibiting seventh-order rotational symmetry. We established that the point group symmetry of these objects corresponds to the dihedral group $D_{7v}$. To gain a deeper understanding of their structure, we performed calculations of the pair radial distribution function of the atoms. This data, along with the diffraction patterns derived from dual quasicrystals, indicates that quasicrystals have a distinct and pronounced diffraction pattern, confirming their crystalline nature and atomic-level orderliness. Furthermore, analysis of the spatial arrangement of atoms from the center to the periphery revealed that the atomic density within these nanoparticles is inhomogeneous. Specifically, there is a noticeable decrease in atomic density as one moves from the center of the crystal towards its periphery.

# INTRODUCTION

Quasicrystals were first experimentally discovered in 1984 by Dan Shechtman [1], who was awarded the Nobel Prize in Chemistry in 2011 for this breakthrough. The discovery of quasicrystals caused a revolution in crystallography, demonstrating that atoms can organize into ordered structures that do not repeat periodically as they do in classical crystals. Quasicrystals possess unique physical and chemical properties, making them promising for various scientific and technological applications.

Let us delve into the analysis of the structural features of quasicrystals and their differences from classical crystals. According to the International Union of Crystallography's Commission on Aperiodic Crystals, a crystal is defined as any solid body that possesses a substantially discrete diffraction pattern. Quasicrystals, in turn, are solid bodies characterized by symmetries forbidden in classical crystallography and the presence of long-range order while also having a discrete diffraction pattern. This definition suggests that quasicrystals are a subclass of the class of crystals. Another subclass includes crystals with rotational symmetry axes of such orders: 1, 2, 3, 4, or 6. Quasicrystals can have symmetries forbidden in classical crystallography, such as seventh-order rotational axes.

The long-range order characterizing quasicrystals differs from that of classical crystals. Since there are different types of long-range order, it is important to define this concept.

Definition: They say that the crystalline structure exhibits long-range order if, wherever a point in space is located, it is possible to specify an algorithm that allows, in a finite number of steps, the determination of whether an element of the crystalline structure (e.g., atom or molecule) is located at that point.

From a group-theoretical perspective, long-range order implies that the crystal's symmetry group must contain an infinite subgroup that determines this order. In ordinary crystals, this subgroup consists of parallel translations, forming a free abelian group whose rank equals the space dimension. A lattice in $n$-dimensional space is defined as the set of points with integer affine coordinates, a free abelian group of rank $n$. In physical Euclidean space $E^n$ ($n = 2$ or $3$), the lattice cannot exhibit fifth or higher sixth-order rotational symmetries, but quasicrystals can possess such symmetries [2].

**Homogeneity of Crystalline Substance [3].** The macroscopic homogeneity of a crystalline substance means that its properties are uniform throughout any section of the material. Regardless of location within a single crystal, an oriented sample of a specific shape and size will exhibit identical physical (optical, mechanical, thermal) and physicochemical properties (e.g., surface solubility, adsorption). Such homogeneity also implies a constant chemical composition and phase state throughout the volume of the material. Statistical studies of macroscopic homogeneity enable the crystalline substance's symmetry to be expressed clearly, aiding in understanding physical properties irrespective of microscopic structure.

A classical crystal is considered a macroscopically homogeneous body with constant properties in terms of chemical and phase composition, as expressed in the uniformity of physical characteristics across any macroscopic volume. However, crystals may exhibit significant differences in properties at boundaries, surfaces, and thin layers.

**Isotropy of a Crystalline Substance.**

The isotropy of a macroscopic crystalline substance, such as glass or polycrystal, means its properties are direction-independent. Ideally, an isotropic crystalline substance would exhibit uniform properties regardless of measurement direction. However, crystalline materials often show anisotropy due to lattice characteristics imposing symmetry that affects their properties.

The symmetry of the crystal lattice determines the degree of anisotropy. For instance, cubic system crystals show minimal anisotropy, while triclinic system crystals display considerable anisotropy. Understanding anisotropy is crucial for the application and research of crystalline materials.

### Heat Transfer in Crystalline Materials.

Heat spreads more effectively in homogeneous solids than in inhomogeneous ones. In homogeneous materials, the thermal conductivity coefficient remains constant throughout, allowing heat to disperse evenly. In heterogeneous materials, phase boundaries and structural differences create thermal resistance, slowing heat propagation. For example, inclusions or grain boundaries can act as barriers to heat transfer.

Homogeneous crystals usually have predictable mechanical properties, such as strength, hardness, and ductility, distributed evenly throughout. In contrast, heterogeneous crystals may show local variations in these properties. Similarly, optical properties like refractive index and transparency are uniform in homogeneous crystals, while non-uniform crystals can exhibit varying optical characteristics.

### Quasicrystals' Unique Properties.

Quasicrystals' quasiperiodic structures lead to complex surfaces with significant variability in chemical composition, impacting their adhesive properties. Quasicrystals often consist of several elements distributed irregularly, resulting in varied atomic compositions and chemical properties across the surface. This variability leads to differences in surface energy and adhesion strength. For instance, areas with high surface energy exhibit stronger adhesion, while areas with low surface energy show weaker adhesion.

The chemical inhomogeneity of quasicrystals creates heterogeneous adsorption centers, affecting reactivity and bonding tendencies. Molecules interacting with quasicrystal surfaces experience varying forces, leading to uneven adhesive behavior. This chemical inhomogeneity can also impact corrosion resistance, with some areas more prone to corrosion altering adhesive properties over time. Aluminum-based quasicrystals, for instance, often have high corrosion resistance, enhancing long-term adhesive properties.

Quasicrystals can be used in coatings to reduce friction, especially in tools and mechanisms where low adhesion and high wear resistance are crucial. Their chemical inhomogeneity also makes them suitable for antibacterial coatings.

In conclusion, the chemical inhomogeneity of quasicrystal surfaces, resulting from their complex quasiperiodic structure, leads to significant variations in adhesive interactions at the microscopic level. These variations impact surface energy, adsorption centers, molecular interactions, and corrosion resistance. The unique properties of quasicrystals make them valuable for various high-tech applications requiring specific adhesive characteristics.

# CONSTRUCTION AND ANALYSIS OF MODEL NANOQUASICRYSTALLINE PARTICLES

The algorithms for constructing quasilattices have been described in my previous works [4-7]. In this study, we extend the approach to constructing quasilattices by employing the matrix-vector representation of the quasicrystalline spider in an orthonormal antiprismatic basis, while also considering the tensor representation of the anisotropic properties of a theoretical crystal. Quasilattices can also be unambiguously reconstructed using quaternions, which provide an efficient and compact means to describe rotations of the basis vector. Additionally, it is worthwhile to note the analogous relationship between the quasicrystalline spider and the local star of the $O$-point in the Delone system [8].

This paper presents the results of studying two models of nanoparticles with seventh-order rotational symmetry. For convenience, we designate the $G1$ model as a nanocrystal consisting of 2,187 atoms and the $G2$ model as a nanocrystal consisting of 78,125 atoms. Both of these nanocrystals have a point group symmetry of $D_{7v}$.

Figures 1 $a$ and 1 $b$ display the atomic model of the $G1$ nanocrystal from two perspectives, where the atoms are represented with a relatively large radius for better visualization of the volumetric structure. It should be noted that if such a quasicrystal exists in reality, it would likely consist of different types of atoms. By drawing planes through all the surface atoms, a polyhedron is obtained, as shown in Figure 1 $c$. This convex polyhedron consists of 28 thin rhombuses with angles of 47.98° and 132.02°, and 14 thick rhombuses with angles of 85.78° and 94.22°. The polyhedron features 44 vertices and 84 edges. According to Euler's formula, $V + F = E + 2$, we have: $44 + 42 = 84 + 2$.

Thus, the geometric characteristics of this polyhedral structure are consistent with Euler's formula.

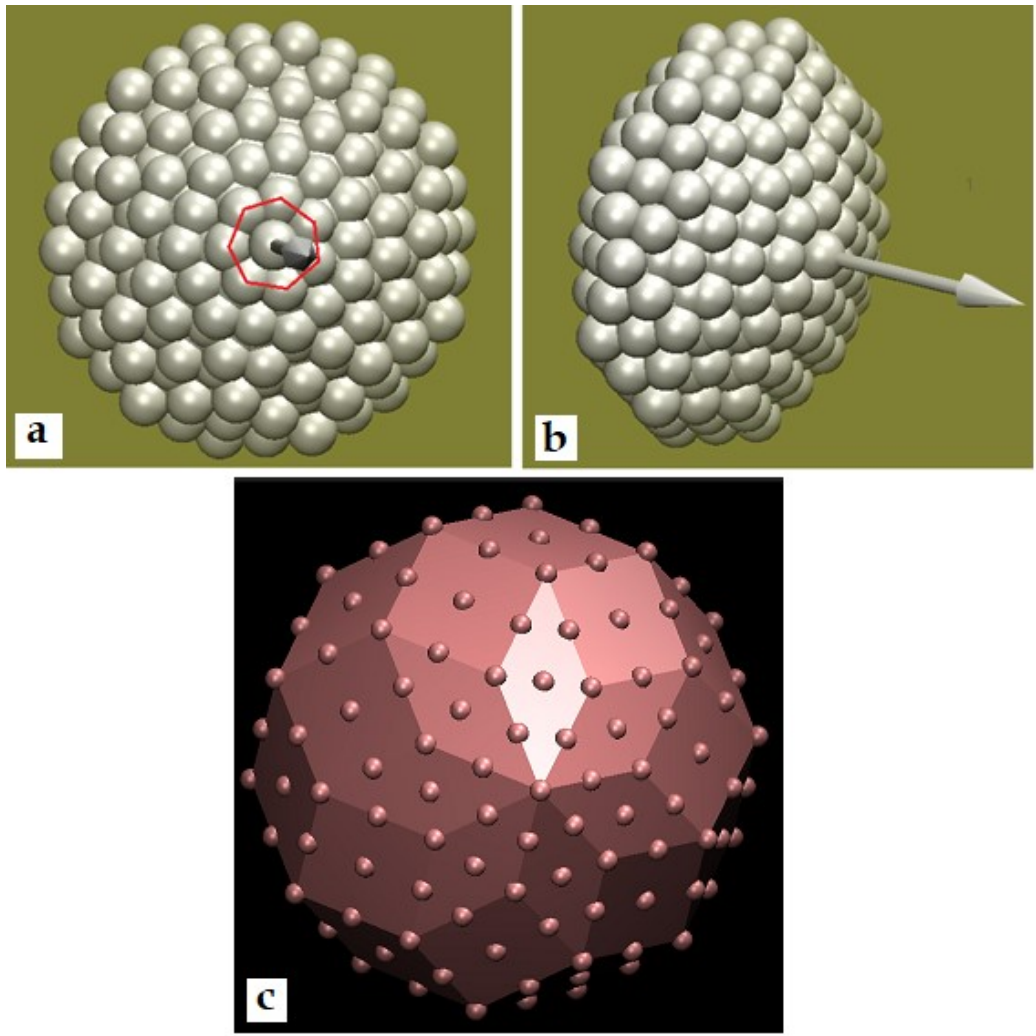

Figure 1. Atomic model of the G1 nanocrystal possessing seventh-order rotational symmetry;

the arrow indicates the direction of this seventh-order rotational axis.

To provide clearer structural insights, Figure 2 presents the G1 nanocrystal from various perspectives, featuring smaller atom sizes compared to Figure 1. In Figures 2 d, 2 e, and 2 f, the crystal orientation allows for the observation of characteristic voids within its central region. Figure 2 j showcases a specific perspective of the nanoparticle, revealing its composition of 15 parallel atomic planes.

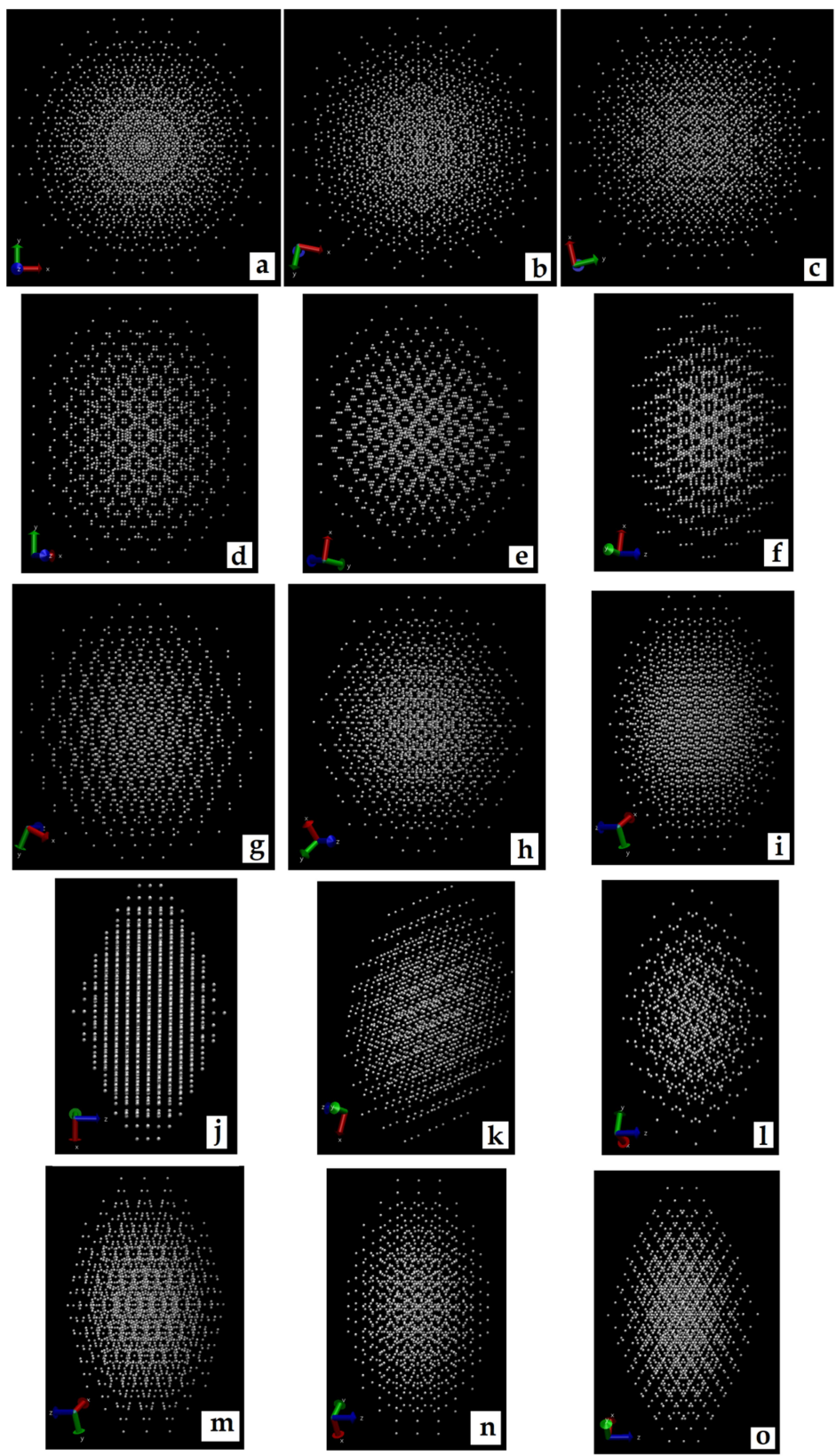

Figure 2. Atomic model of the G1 nanocrystal shown from different perspectives.

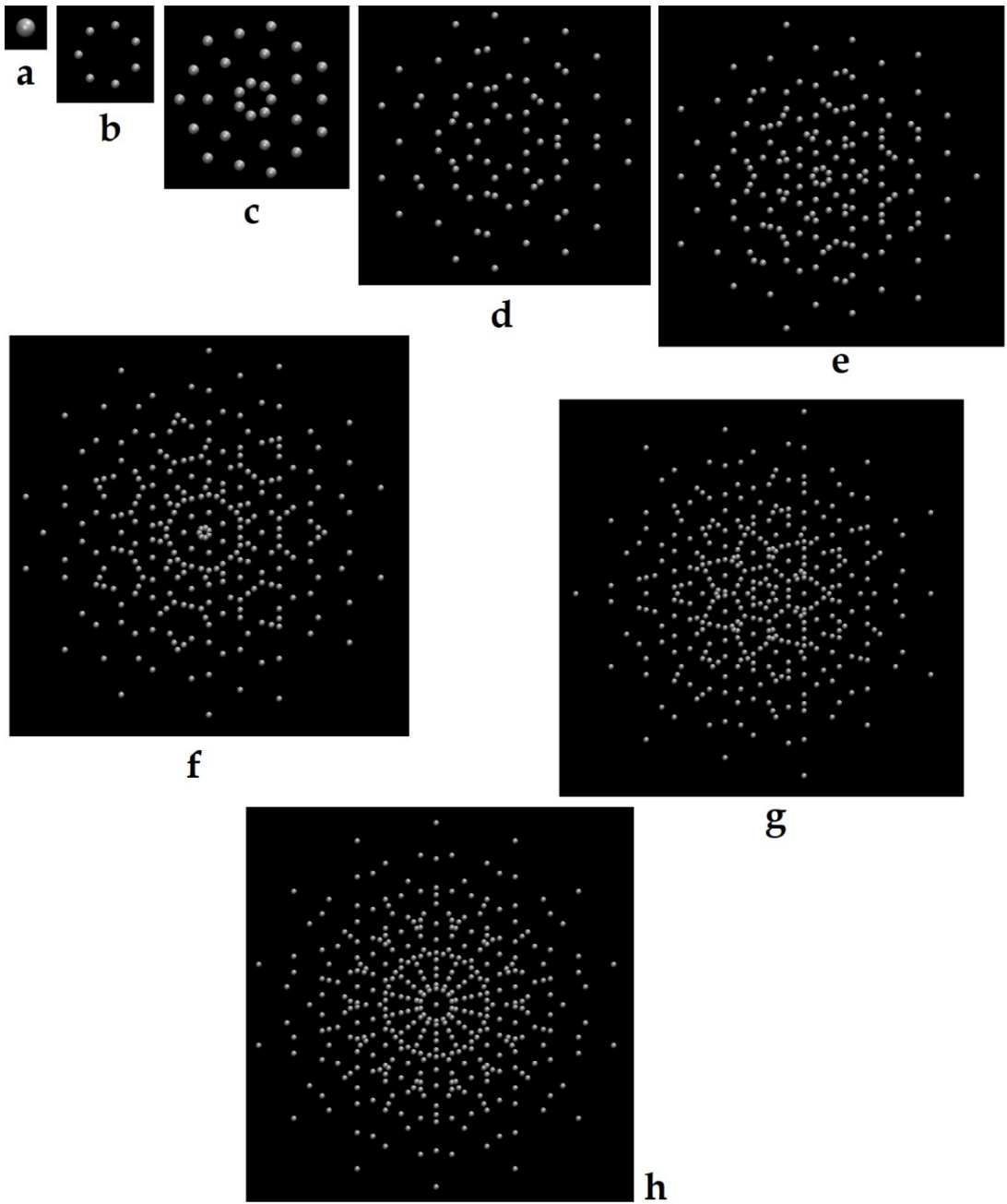

Figure 3. Image of sequential atomic planes; *a* - represents the vertex atom on the crystal surface, through which the seventh-order axis passes; *h* - indicates a plane that passes through the center of the crystal, perpendicular to the seventh-order rotational axis.

If we take a crystal and cut it along parallel atomic planes, we obtain a set of planes. In classical crystals, the arrangement of atoms remains consistent from plane to plane. That is, if we take any two planes from the set of parallel planes of a layered classical crystal, the atoms will form an identical crystalline pattern (excluding the surface atoms). For example, if we take a *NaCl* crystal, which has a face-centered cubic lattice, and layer it along planes of type (1, 0, 0), we will observe planes with atoms positioned at the vertices of squares and at their centers. Any two parallel planes, regardless of their size and ignoring surface atoms, will be equivalent.

In contrast, if we perform a similar layering in quasicrystals, each parallel atomic plane will be populated by atoms in a unique arrangement. Figure 3 illustrates the sequential layers of parallel atomic planes that constitute the *G*1 crystal. It depicts the layering of the crystal on one side of the plane in Figure 3 *h*. On the opposite side of the plane in Figure 3 *h*, the planes are identical in atomic arrangement but rotated by an angle of $2\pi/14$.

In Figures 3 *a*, 3 *b*, 3 *c*, 3 *d*, 3 *e*, 3 *f*, 3 *g*, and 3 *h*, the number of atoms is 1, 7, 28, 77, 161, 266, and 357 respectively. Notably, the number of atoms in Figures 3 *b* through 3 *g* is divisible by 7. However, in Figures 3 *a* and 3 *h*, the number of atoms is not divisible by 7. Considering that the number 357 can be expressed as 326 + 1, where 1 represents the central atom, the number 326 is divisible by 7. The total number of atoms in the *G*1 model is 2,187, which is $3^7$. While 2,187 is not divisible by 7, if we subtract 3 (which represents the central atom and the two surface atoms through which the seventh-order rotational symmetry axis passes), we get 2,184. Dividing 2,184 by 7 yields 312. Thus, the number of atoms excluding the central atom and two surface atoms is divisible by 7.

It should be noted that the sets of atoms depicted in Figures 3 *b* to 3 *g* exhibit seventh-order rotational symmetry. The set of atoms shown in

Figure 3 *h* exhibits fourteenth-order rotational symmetry. Details of this remarkable fact will be discussed further.

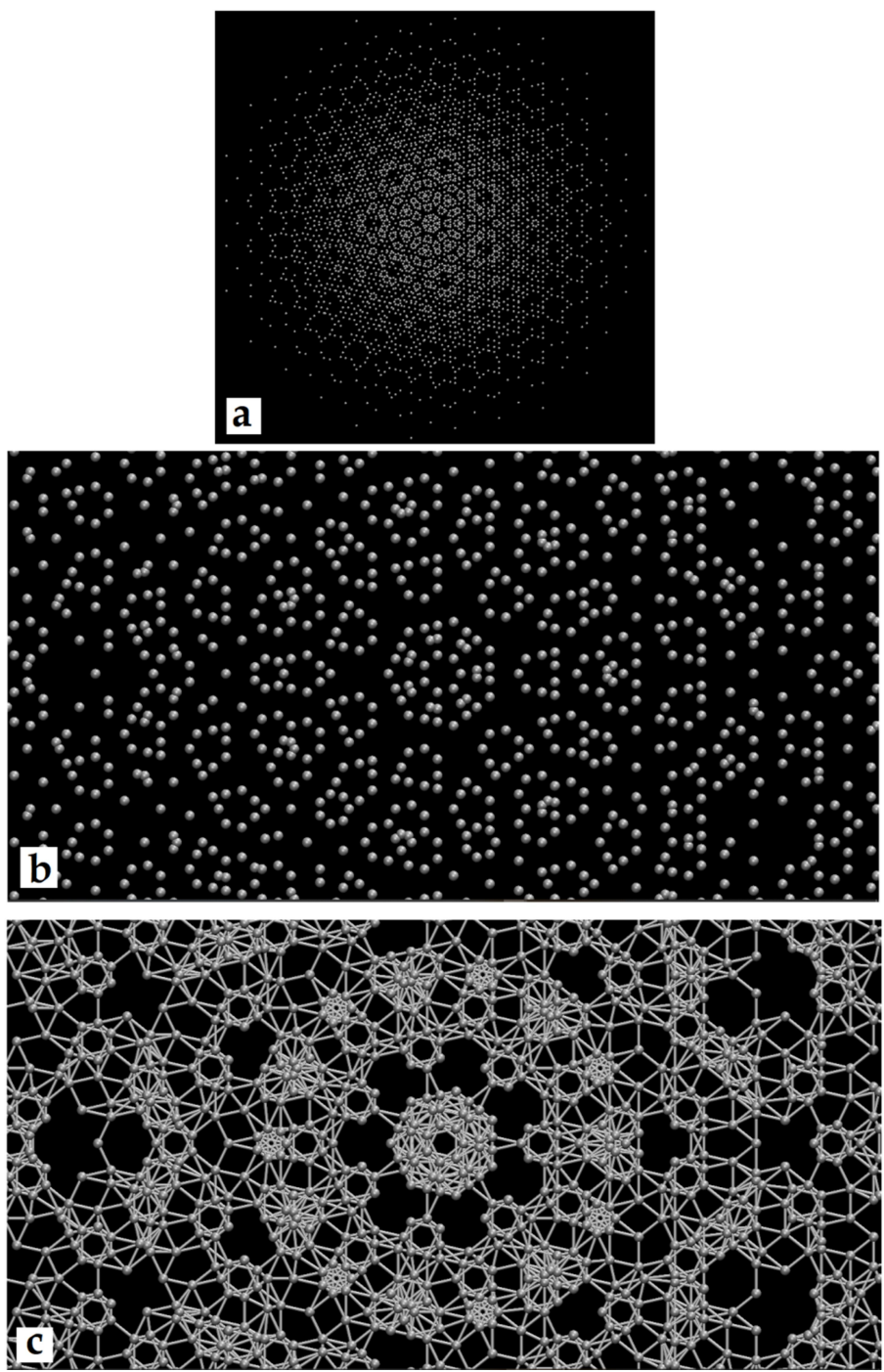

Figure 4. *a* - quasicrystalline ensemble consisting of 3,535 atoms;
*b* - an enlarged view of the central region of ensemble *a*;
*c* - connected atoms from ensemble *b* with segments of a certain length.

In Figure 4 *a*, the fifth atomic layer is shown, counting from the central plane of crystal G2. The entire crystal G2 consists of 29 parallel atomic planes, similar to the 15 parallel atomic planes of crystal G1 depicted in Figure 2 *j*. Figure 4 *b* shows an enlarged central area from Figure 4 *a*. In Figure 4 *c*, each atom is connected by segments of a certain length to its nearest atoms, allowing for a corresponding quasiperiodic tiling of the plane.

The total number of atoms in the G2 model is 78,125, and the number 78,125 is not divisible by seven (78,125 % 7 = 5). In the crystal model G2, in addition to three atoms, one at the geometric center of the crystal, two atoms at the vertices of the approximation polyhedron, there are also two atoms located in the centers of the seventh planes of layering on both sides of the central plane. If you take a characteristically oriented crystal and orthogonally project all its atoms onto a plane, you get a certain set. Figure 5 *a* shows such a crystalline set, possessing 14-fold rotational symmetry. Note that the G1 crystal itself does not coincide with itself when rotated by an angle of $2\pi/14$. However, its orthogonal projection of atoms coincides with itself when rotated by an angle of $2\pi/14$ relative to the central atom in this set. This fact illustrates an interesting and important property of crystal symmetry. In three-dimensional space, the crystal itself may not possess a certain rotational symmetry, whereas its two-dimensional projection may demonstrate such symmetry. This is because projecting atoms onto a plane simplifies the crystal's geometry. In other words, such projections lose information, such as whether the symmetry axis in the crystal along which the imaging was performed is rotational or mirror-rotational/inversional-rotational. Consider, for example, the symmetry group of the icosahedron, which consists of: 1, $12\,C_5$, $12\,C_5^2$, $20\,C_3$, $15\,C_2$, $i$, $12\,S_{10}$, $12\,S_{10}^3$, $20\,S_6$, $15\,\sigma$. In work [1], an electron diffraction pattern is given, from which it follows that the pattern itself possesses 10-fold rotational symmetry, although the icosahedron only has $S_{10}$ axes.

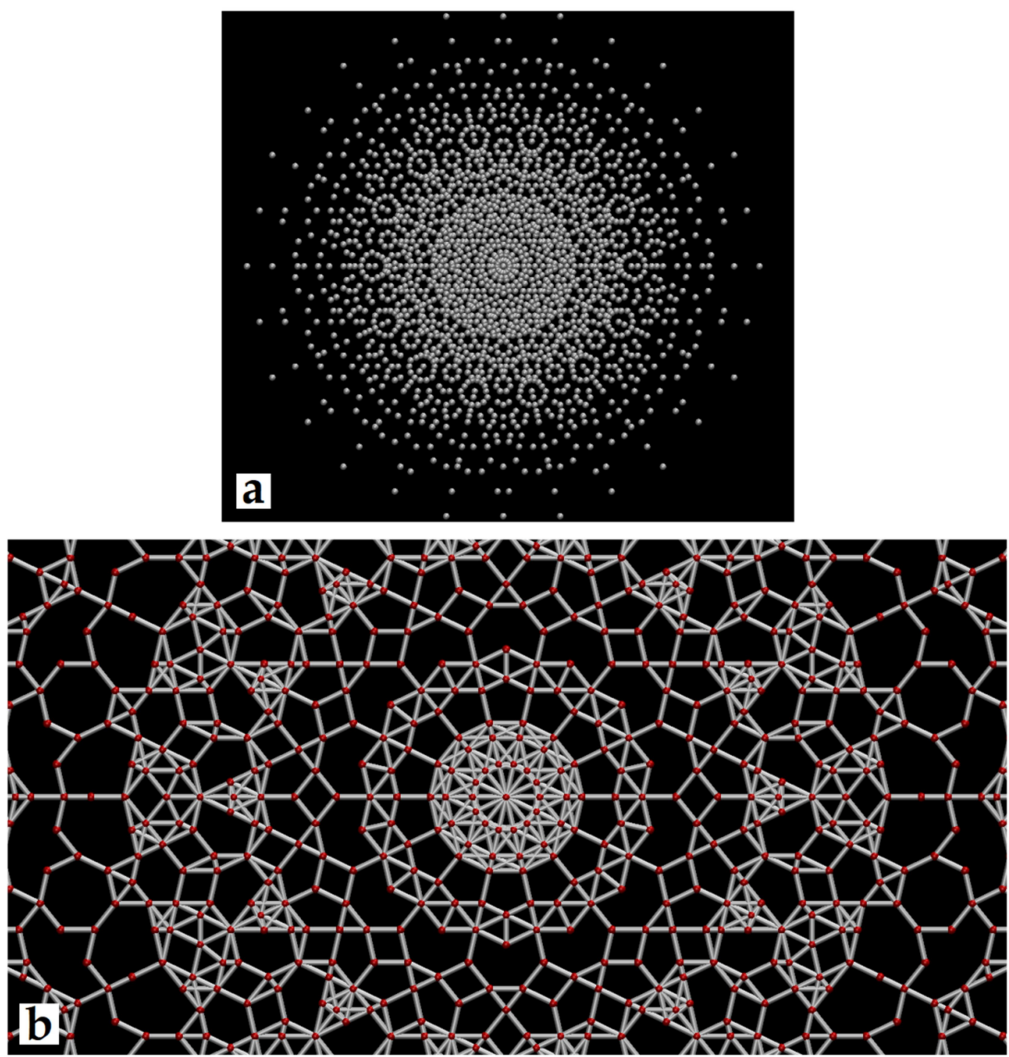

Figure 5. *a* – orthogonal projection of the atoms in the *G*1 model onto a plane perpendicular to the 7-fold rotational symmetry axis (i.e., the *z*-coordinate of each atom is 0); *b* – the atoms from *a* are connected by line segments of a certain length.

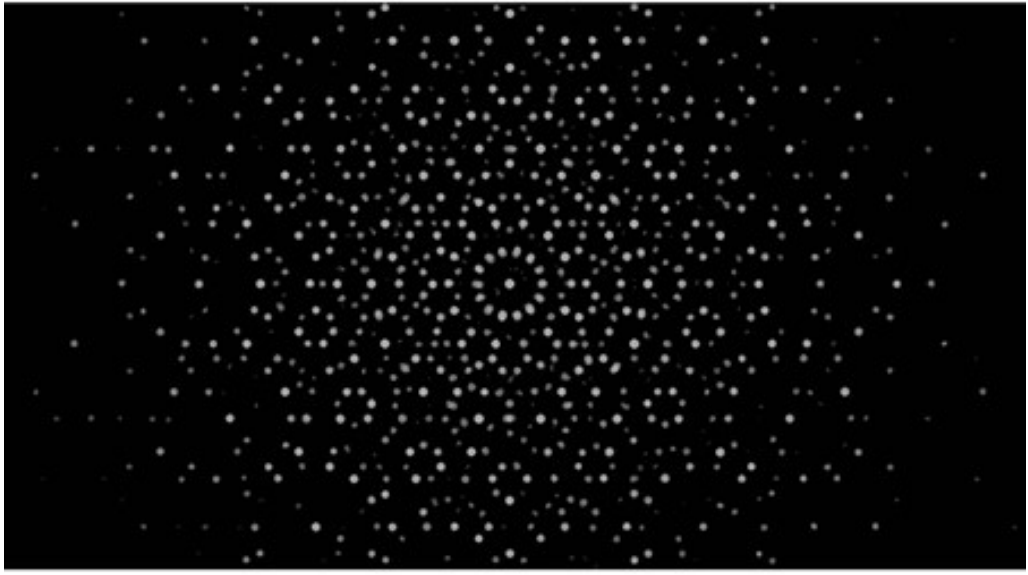

Figure 6. Diffraction pattern obtained from the dual crystal to the G1 crystal.

Figure 6 shows the diffraction pattern obtained from the crystal dual to the G1 crystal. This pattern indicates the presence of 14-fold rotational symmetry in the crystal, even though the crystal itself does not coincide with itself when rotated by an angle of $2\pi/14$. The characteristic sharp peaks on the pair radial distribution function of atoms in Figure 7 also indicate the crystallinity of the studied G1 nanoparticle. The function shown in Figure 7, or more precisely the relative positioning of the peaks, will be useful in identifying experimentally found crystal phases similar to G1, possessing 7-fold rotational symmetry.

Analysis of the spatial arrangement of atoms from the center to the periphery of the crystals G1 and G2 showed that the atomic density in these nanoparticles is non-uniform. Specifically, there is a decrease in atomic density as the distance from the center of the crystal to its periphery increases. This phenomenon can be explained by the features of crystal formation and growth, where certain types of atoms occupy denser positions in the central regions, while peripheral zones remain less populated. This structure may arise due to differences in the sizes and masses of atoms: intuitively, it seems that less massive atoms tend to concentrate in the central regions of the crystal, while more massive atoms are located on the periphery of the crystal. However, it could be exactly the opposite.

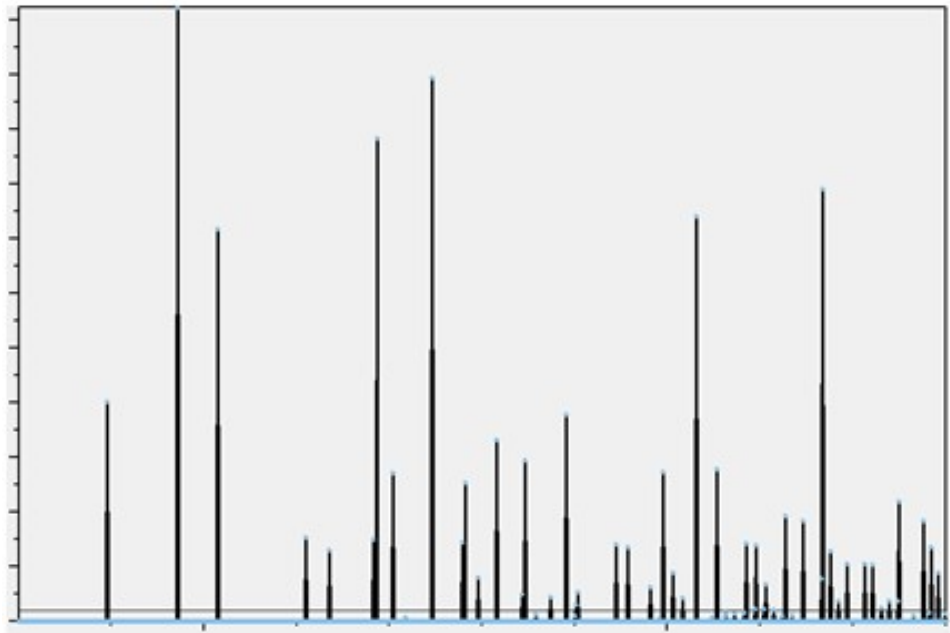

Figure 7. Pair radial distribution function of atoms calculated for the G1 crystal.

Earlier, differences between homogeneous and non-homogeneous materials were listed. These differences are related to the fact that, in particular, in classical crystals, inhomogeneities are randomly arranged. In principle, a crystal can be grown homogeneously. Defects in the structure can arise due to certain crystallization conditions. However, in monoquasicrystals, inhomogeneities are not random but follow a strict law. Therefore, quasicrystalline materials should not be considered as non-homogeneous materials in the classical sense. They have a systematically mathematical inhomogeneity. Consequently, quasicrystalline materials have a discrete nature of anisotropic-inhomogeneous properties.

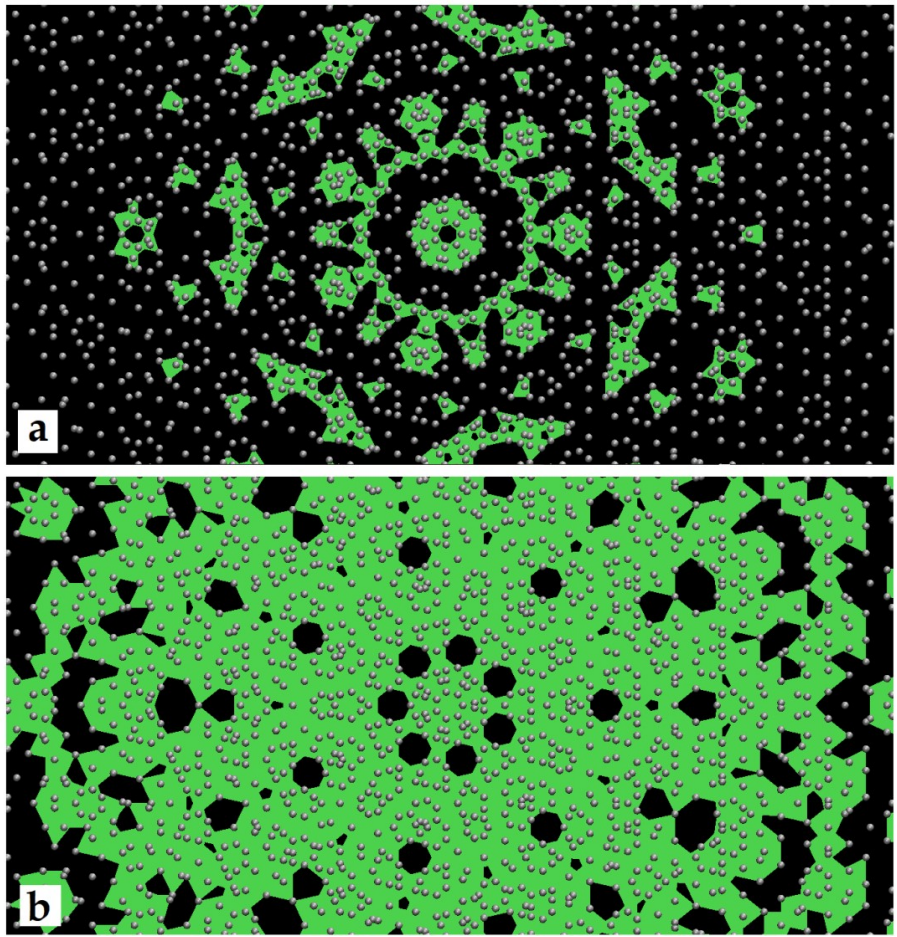

Figure 8. Map of local inhomogeneities;
quasicrystalline plane consisting of 3535 atoms;
this is the fifth layer from the central plane of the G2 crystal.

Figure 8 shows the homogeneity map on the atomic plane of the G2 crystal. In Figure 8 *a*, green areas highlight regions with maximum local atomic density. In Figure 8 *b*, the criterion for filling with green color captures a larger range in local density, but black areas still indicate local 'voids' in the crystal.

# CONCLUSIONS

This study shows that quasicrystals have a radial density gradient, indicating complex self-organization mechanisms during the synthesis of nanoparticles, which can be important for their physical and chemical properties. Furthermore, the 7-fold symmetry axis and the discovered inhomogeneity of atomic density suggest that these nanoparticles may possess unique physical properties, such as anisotropy in optical and magnetic characteristics. Such properties can be useful in various applications, including the creation of new materials with desired properties, the development of sensors and catalysts, as well as in nanotechnology and biomedical research.

The inhomogeneity of atomic density and the unique symmetries of quasicrystals can lead to specific physical properties, such as high hardness, low thermal conductivity, and unique electrical properties. These characteristics make quasicrystals attractive for use in new materials and technologies. For example, they can be used as components in thermoelectric materials, which convert heat into electricity or vice versa. Additionally, quasicrystals may find applications in creating new types of coatings with high resistance to wear and corrosion.

Another interesting feature of quasicrystals is their potential in biomedical applications. Due to their unique structure and properties, quasicrystals can be used to create new types of biocompatible materials, which can be applied in implants and prosthetics. Moreover, their unusual optical properties can be useful in developing new types of sensors and diagnostic tools.

Thus, quasicrystals represent an interesting and promising class of materials, combining unique structural and consequently physical-chemical characteristics. Their study and the development of new applications open wide opportunities for innovation in various fields of science and technology. It is important to continue research in this area to better understand the mechanisms of quasicrystal formation and growth, as well as to identify all potential applications of these unique materials. The obtained results not only expand fundamental knowledge about the structure and properties of quasicrystalline nanoparticles but also open new opportunities for their practical use.